# Hybrid modelling of a sugar boiling process

## P. Lauret[*], H. Boyer and J.C. Gatina

Laboratoire de Génie Industriel, Université de La Réunion. 15, avenue René Cassin – BP 7151
97715 Saint Denis Messageries Cedex 9.

**Abstract:** The first and maybe the most important step in designing a model-based predictive controller is to develop a model that is as accurate as possible and that is valid under a wide range of operating conditions. The sugar boiling process is a strongly nonlinear and nonstationary process. The main process nonlinearities are represented by the crystal growth rate. This paper addresses the development of the crystal growth rate model according to two approaches. The first approach is classical and consists of determining the parameters of the empirical expressions of the growth rate through the use of a nonlinear programming optimization technique. The second is a novel modeling strategy that combines an artificial neural network (ANN) as an approximator of the growth rate with prior knowledge represented by the mass balance of sucrose crystals. The first results show that the first type of model performs local fitting while the second offers a greater flexibility. The two models were developed with industrial data collected from a 60 m$^3$ batch evaporative crystallizer.

## 1. Introduction

In order to optimize the performance of the sugar boiling process, a model-based predictive control (MBPC) is investigated. The major part of the MBPC design effort must be devoted to modeling (Richalet, 1993). In other words, the key of a successful implementation of this control method is to develop a model that is as accurate as possible for all the conditions prevailing in the process.

The sugar crystallization process has time-varying dynamics and it exhibits strong nonlinear effects. The main process nonlinearities are included in the growth rate. Such nonlinear and nonstationary processes present challenges to process control (Joshi et al., 1997).

Several attempts have been made to control this process. Feyo de Azevedo et al (1989), simulated the operation of a sugar batch vacuum evaporative crystallizer with a generalized predictive control (GPC) architecture. Despite the effectiveness of the GPC approach, the main difficulty in the implementation came from the necessity to define two linear models in order to cope with two different feed-stocks. Qi & Corripio (1985) implemented a modified version of a dynamic matrix control (DMC®) to control a vacuum pan. As the standard DMC linear predictive model had difficulty in handling the nonlinear effect, a nonlinear model of the process was introduced into the control scheme to predict the response of the controlled variables. Nevertheless, the adjustments on the manipulated variables were still made using the standard DMC approach. In their conclusions, the authors suggest the need to determine several sets of numerical coefficients for the step response in order to deal with the dynamic nonlinear process characteristics.

These contributions showed that efficient control strategies can be penalized by models that are not valid over a range of operating conditions.

The sugar boiling process must handle different feedstocks (syrup of different purity) and must produce several type of sugars. So, in order to avoid the development of separate black-box linear models for each operating point, which can be somewhat time consuming, a nonlinear model, applicable to all conditions, is needed .
As the main nonlinearities are included in the sucrose growth rate, it will be the key parameter of the proposed model .
The crystallization phenomenon is complex, so the development of a pure white (or totally first-principles) nonlinear growth rate model is prohibited.
In other words, the model of the growth rate will contain some parameters fitted to data collected from the process. Georkagis (1995) termed these types of model *grey models*. The greyness of the model denotes the fact that these models are a combination of data-driven and knowledge-driven components.

[*] Corresponding author. Fax: +262 93 86 65; e-mail: lauret@univ-reunion.fr

This paper describes two types of models of the growth rate, each with its own specific shade of grey. The first one consists of estimating the kinetic parameters of empirical expressions through the use of a nonlinear programming (NLP) optimization technique. The second one proposes a more flexible empiricism through the use of an artificial neural networks (ANN) to model the growth rate (Rawlings et al., 1993). Here, the ANN is the data-component that is combined with a knowledge-driven one through the use of a material balance. This combination gives a grey model. Psichogios & Ungar (1992) called them *hybrid neural network* models.

The development of these models were based on experiments carried out on a large-scale sugar-batch evaporative crystallizer.

The main objective of this paper is to demonstrate that NLP performs local fitting while the hybrid neural network gives rise to a more flexible model of the growth rate. Consequently, the scope of nonlinearities of concern is much wider.

The paper is organized as follows: Section 2 describes the experimental set-up and the process operation of sugar boiling in batch vacuum pans. Section 3 discusses the sugar boiling process model while Section 4 focuses on the growth rate model. Section 5 deals with the first type of growth rate model. Two parameterized models are investigated. It will be shown that, while it is difficult to find the proper parameterization, these models have poor predictive capability outside the region where the kinetic parameters are fitted. Section 6 discusses the second type of model with a special emphasis on the specific training procedure of the hybrid ANN. It will be demonstrated that such a model performs better under varying conditions and with less modeling effort. Section 7 gives the conclusions.

## 2. Process operation and experimental set-up

Usually, the crystallization of sugar is carried out by boiling sugar liquor in a batch-type evaporative crystallizer called a vacuum-pan in the sugar industry. Such a vacuum-pan is represented in Figure 1.

It is a single-stage vertical evaporator heated by steam condensing inside a calandria. The vapor from the pan is condensed by direct contact with cooling water in a system that also provides the vacuum control.

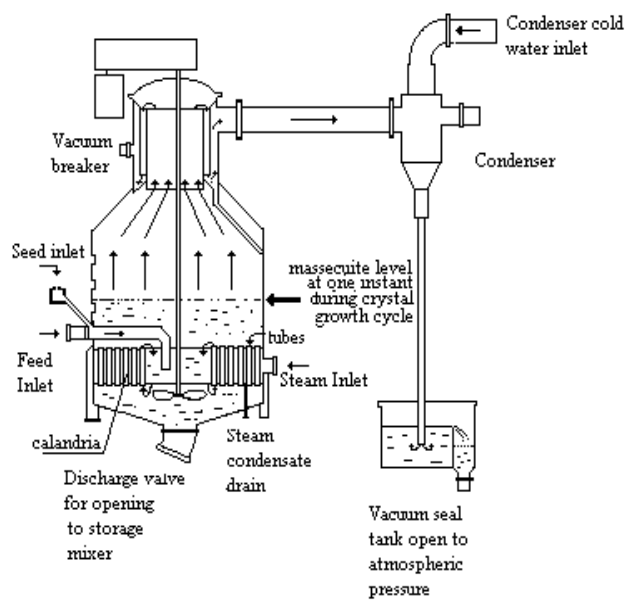

Fig. 1. Sugar batch vacuum pan.

The basic steps involved in any crystallization process are (Tavare, 1995): achievement of supersaturation, formation of crystal nuclei (nucleation) and subsequent growth to form large crystals. Supersaturation, generated here by evaporation of water, is the driving force for crystallization. In the present case, it is expressed as the ratio of concentration of sucrose in solution to its saturation concentration for the same process conditions. Thus, crystal growth can only occur when supersaturation exceeds the saturation state. In the sugar industry, it is common practice to avoid nucleation. Finely powdered sugar or seed grain is introduced into the pan to serve as nuclei that grow into product-size sugar crystals. All along the batch, the supersaturation must be kept within safe limits in order to avoid the formation of new crystals or the dissolution of the existing ones. This safe zone, where crystals can only grow, is called the metastable zone.

The batch cycle is subdivided into several sequential phases. Table 1 lists the more important ones.

Table 1. Phases of the sugar boiling process

| # | phase |
|---|-------|
| 1. | initial filling |
| 2. | concentration |
| 3. | seeding |
| 4. | boiling (or growing the grain) |
| 5. | discharging to a storage mixer before centrifugation |

First, a certain amount of sugar liquor is introduced into the pan. The liquor is concentrated by evaporation under vacuum until the supersaturation reaches a predetermined value. At this point, seed crystals are introduced to induce the production of crystals. To ensure the growth of crystals, supersaturation is maintained by evaporating water and by feeding the pan with fresh syrup or molasses (juice with less purity). At the end of a batch, the pan is filled with a suspension of sugar crystals in heavy syrup which is dropped into a storage mixer before centrifugation. This suspension is called the massecuite. The syrup around the crystals is called the mother liquor. Each panful of massecuite is called a strike. More details about the sugar boiling process can be found in (Chen, 1985).

The industrial unit at Bois Rouge Sugar Mill is a fixed-calandria type, specially designed with large diameter short tubes and a large downtake for the circulation of the heavy viscous massecuite. It is equipped with a propeller stirrer and has a working volume of 60 m$^3$. Feed syrup enters near the pan central downtake, mixing with the bulk of the massecuite, which then circulates upwards through the calandria tubes. The physical measurements available on the industrial unit are listed in Table 2.

Table 2. Instrumentation of the industrial unit

| # | physical measurements |
|---|---|
| 1. | massecuite temperature |
| 2. | level of massecuite in the pan |
| 3. | brix of mother liquor by means of a process refractometer. |
| 4. | brix of massecuite by means of a gamma density meter (radiation gauge) |
| 5. | electrical conductivity of the massecuite |
| 6. | intensity of the stirrer current |
| 7. | vacuum pressure inside the pan |
| 8. | steam pressure |
| 9. | flow rate of feed syrup |

The brix of a solution is the concentration of total dissolved solids (sucrose plus impurities) in the solution. Purity is the proportion of sucrose in brix. Present pan control involves the manipulation of feed rate in order to follow a pre-set profile of brix. The development of real time, multitasking software under LINUX (a free UNIX-like operating system) provided complete interaction with the control and supervision system of the industrial unit (MODUMAT® by Bailey). The performance of the system allows a scanrate of 10 measurements per second.

This interaction was not only restricted to the industrial acquisition of physical measurements but has opened the way to the development of software sensors. A software sensor of supersaturation was developed (Lauret et al., 1997). This tool also gave real time estimates of the purity of the mother liquor and crystal content in the pan. The crystal content is the solid fraction in the massecuite.

The effectiveness of this software tool was verified during the last sugar campaign at Bois Rouge Sugar Mill.

These estimates together with the physical measurements were recorded in large databases. These databases, each related to a batch, formed the basis for the development of the two types of growth rate models under MATLAB®. Table 3 shows the fields of a database .

Table 3. Record of a database

| # | Notation | Field |
|---|---|---|
| 1 | Time | time of observation |
| 2 | Phase | # Phase ( concentration, seeding ,…) |
| 3 | Bxml | Brix of mother liquor |
| 4 | Bxmc | Brix of massecuite |
| 5 | $\gamma$ | Electrical conductivity of massecuite |
| 6 | L | level in the pan |
| 7 | T | Temperature of the massecuite |
| 8 | I | Intensity of stirrer current |
| 9 | Vac | Vacuum in the pan |
| 10 | Ps | Steam pressure |
| 11 | Pfs | Purity of feed syrup |
| 12 | CC | Crystal Content |
| 13 | Pml | Purity of the mother liquor in the pan |
| 14 | SS | Supersaturation |
| 15 | C | Weight of crystals in the pan |

The weight of crystals was inferred from the level, the brix of massecuite measurements and the crystal content. A batch or a strike lasts approximately 4.5 hours and contains 16466 of such records in the database. Some of these observations enabled the development of the models. Nonetheless, it is worthwhile noting that these databases serve also for simulating different control schemes.

## 3. The sugar boiling process model

The complete nonlinear dynamic model of the sugar boiling process is described in the appendix. The model relies on classical material and energy balances. In the case of particulate systems such as crystallization processes, a third balance is introduced in order to characterize the population of crystals. Solution of this population balance yields the crystal size distribution. The latter, through its information aggregates such as the mean size or the coefficient of variation, gives an idea of the product quality. The model is then completed with the rate terms of heat and mass transfer, and the thermodynamic equilibrium relationships such as the sucrose solubility. The balances are generally

"true" e.g. they are not affected by the application to the model of syrups of varying purity. Other relationships including growth, solubility are quite dependent on these conditions (Wright & White, 1974).

In addition, the process exhibits time-varying dynamics as the conditions in the pan, particularly viscosity, evolve during the strike. This nonstationary character is highlighted, for instance, by the overall heat transfer coefficient.

The key parameter in optimizing the pan performance is the growth rate. Effectively, this parameter directly influences the pan's performance in terms of the production rate and product quality, as it is coupled to the population balance.

## 4. The growth rate model

The crystal growth phenomenon is complex because of a large number of interacting variables. Furthermore, the effect of some of these, such as impurities, on the kinetics is nonlinear or even unpredictable. A large number of models that include these variables in complex nonlinear expressions, have been proposed to describe this parameter (Gros, 1979; Wright & White, 1974 ; Feyo de Azevedo et al., 1989; Guimarães et al., 1995 ).
Modeling difficulties arise from poorly understood phenomena. Consequently, application of MBPC can be hampered by this sensitive model parameter.

### 4.1. The mass balance on sucrose crystals

As the focus is on the growth rate model, the relation of interest is the material balance on sucrose crystals.

$$\frac{dC}{dt} = R_G \cdot A_T \qquad (1)$$

In this equation, $R_G$ is the growth rate while $A_T$ is the total surface area of crystals. With the assumption that the number of crystals remains constant all along the strike, the total surface area can be expressed according to the weight of crystals C.

$$A_T = \frac{k_a \cdot N^{1/3}}{(\rho \cdot k_v)^{2/3}} \cdot C^{2/3}$$

(2)
where:

$k_a$ is the surface shape factor of a crystal

$k_v$ is the volume shape factor

$\rho$ is the density of sucrose crystals

N is the number of crystals

C is the mass of crystals in the pan

The values of these parameters are supposed to be known and taken from literature. The number of seed crystals is determined from the sugar plant laboratory. Table 4 gives the values of these data.

Table 4. Values of known parameters of the model

| $k_a$ | $k_v$ | $\rho$ | N |
|-------|-------|--------|---|
| 5.02 | 0.75 | 1580 | $4e^{11}$ |

### 4.2. The experimental observations

In the case of sugar crystallization, the process variables that affect the crystal growth rate are rather well known. Nevertheless, the dependence of these parameters and their nonlinear effect on the growth rate makes modeling difficult. These variables are mainly supersaturation, temperature, dissolved impurities and massecuite viscosity. The latter depends on the solid fraction in the massecuite.
For the present industrial set-up, the experimental observations available in assessing and validating the growth rate were:

SS      Supersaturation
CC      Crystal content
Pml     Purity of mother liquor
C        Mass of crystals in the pan

Crystal content is the solid fraction in the massecuite, while purity is the proportion of sucrose in total dissolved solids of mother liquor. Temperature was not taken into account here because its effect was not significant in the region of interest. This latter phase was obviously the so-called growing the grain phase.
These observations were extracted from the databases. Four experimental strikes were chosen for illustrating the results. These strikes are numbered strikes 1 to 4.
Figure 2 shows the profiles of supersaturation and mass of crystals for strike 1.

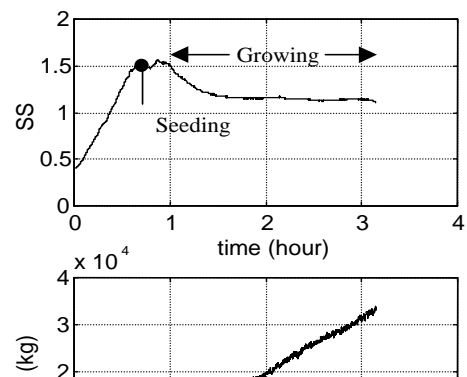

Fig. 2. Profiles of supersaturation and mass of crystals (strike 1)

The growth rate $R_G$ will be modeled according the two approaches described below.

## 5. Crystal growth rate model through NLP optimization technique

The mass balance on sugar crystals (Eq. 1) is a nonlinear first-order differential equation. Its solution represents the evolution of the mass of crystals within the pan during the strike. This variation is a function of the time and kinetic parameters. These parameters can be determined by matching the experimental observation $C_{obs}$ with the solution of Eq. 1, e.g. the model response $C_{mod}$. This results in a nonlinear parameter estimation problem.

### 5.1. Kinetic correlations

For the kinetic correlations, convenient working relations for pragmatic situations are proposed in the literature (Tavare, 1995). In the present case, two parameterized models were investigated. At first sight, it might have been plausible to take into account the ratio $\dfrac{100 - CC}{CC}$ of slurry voidage to solid fraction in this large scale process. So, the following model structure was investigated.

$$R_G = k_g (SS-1)^g \left( \frac{100 - CC}{CC} \right)^e \exp[\alpha(1-Pml)] \quad (3)$$

Table 5. Features of the NLP technique

| Observations | model response | Parameters |
|---|---|---|
| $C_{obs}$, $SS_{obs}$, $CC_{obs}$, $Pml_{obs}$ | $C_{mod}$ | $\Theta = [k_g, g, e, \alpha]$ |

Table 6. Results for the first empirical expression. Case of strike 1 only

Table 5 lists the features of the nonlinear estimation problem.

A variant is introduced here in the parameter estimation scheme because Eq. 1 must be integrated in order to obtain the model response $C_{mod}$. At each iteration, the mass balance is integrated with the parameter estimates of the previous iteration. The cost function J can be then calculated. Figure 3 illustrates the methodology.

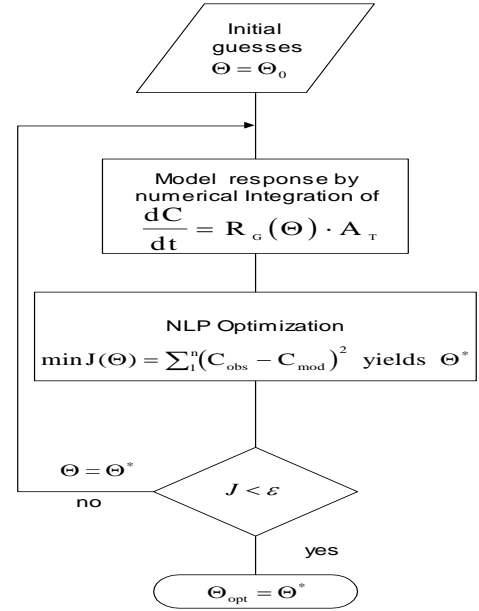

Fig. 3. Estimation of kinetic parameters

The numerical integration was made by using a fourth-order Runge Kutta method. The optimization algorithm was a Gauss Newton method. Both of them were available under the MATLAB® environment. The guesses on initial values were made according to typical values of crystallization parameter estimates provided by (Tavare, 1995). Table 6 shows the results for the strike 1.

| | Initial guesses | Parameters found |
|---|---|---|
| # | $\Theta_0$ | $\Theta^*$ |

| 1 | [ $1\times10^{-4}$ 1.0 1.00 -0.1 ] |
|---|---|
| 2 | [ $1\times10^{-4}$ 1.5 1.00 -0.1 ] |
| 3 | [ $1\times10^{-4}$ 2.0 1.00 -0.1 ] |
| 4 | [ $2\times10^{-4}$ 1.3 0.05 -4.0 ] |
| 5 | [ $1\times10^{-4}$ 3.0 1.00 -0.1 ] |

| [ $0.798\times10^{-4}$ 1.184 0.313 +1.604 ] |
|---|
| [ $2.174\times10^{-4}$ 1.290 0.036 -3.795 ] |
| [ $1.477\times10^{-4}$ 1.331 0.135 -0.087 ] |
| [ $2.679\times10^{-4}$ 1.260 -0.021 -5.756 ] |
| [ $8.024\times10^{-4}$ 3.019 1.003 +1.54 ] |

| 3 | $1.377\pm0.386 \times10^{-4}$ | $0.914\pm0.152$ |
|---|---|---|
| 4 | $6.440\pm4.650\times10^{-4}$ | $1.847\pm0.302$ |

The following observations can be made:
First, the convergence of the algorithm was very slow and difficult. Second, the procedure exhibits a strong sensitivity to the initial guesses of the parameters. Moreover, the results seem not to be reasonable for line 1 and 5 of Table 6. For line 1, the parameter $\alpha$ must be negative since as purity decreases so does the rate of crystallization. For line 5, the order g must be less than 3 (Tavare, 1995). For the latter, the optimization algorithm has reached a local minimum. It seems that there are too many parameters for the optimization algorithm. So, other optimization algorithms such as Simplex Nelder-Mead and Levenberg-Marquardt were tested. But, the results were not really good and were different for each algorithm. In all cases, the convergence to local optima may have occurred and the estimates of the parameters appear to be correlated. With this large number of adjustable parameters, extensive experimentation may be required in order to find the best parameterization suitable for the process at hand.

Due to these modeling difficulties, a classical empirical expression for the growth rate was used where all the other variables are lumped in the overall growth rate constant $K_G$:

$$R_G = K_G \cdot (SS - 1)^g \qquad (4)$$

This model structure is the one that is usually used in the crystallization engineering practice. Table 7 shows the results.

Table 7. Results for the second expression

| strike | $K_G$ | g |
|---|---|---|
| 1 | $2.198\pm0.090 \times10^{-4}$ | $1.508\pm0.008$ |
| 2 | $3.024\pm0.806 \times10^{-4}$ | $2.209\pm0.195$ |

Fortunately here, whatever the initial guesses for each strike were, the procedure converges to the same values of the two parameters. Moreover, all the optimization algorithms tested led to the same results. So, the NLP technique works better with this simplified model structure. Furthermore, the results are consistent with previous laboratory work (Guimarães et al.,1995).

But, despite the study in parameter uncertainty, the scatter in parameter estimates is significant.
The following could be reasons for this.
First, it is important to note that at present time, supersaturation is not a controlled variable in the industrial control scheme. So, the supersaturation profiles vary between strikes (see Figure 4). Obviously, this variable strongly affects the result of the optimization problem. The order g of the growth rate is particularly dependent on the supersaturation. It can be said that the fit to the model depends heavily on the local operating conditions. Nevertheless, the complexity of the phenomena that occurs in the pan makes any attempt to clearly explain this scattering in parameters difficult.

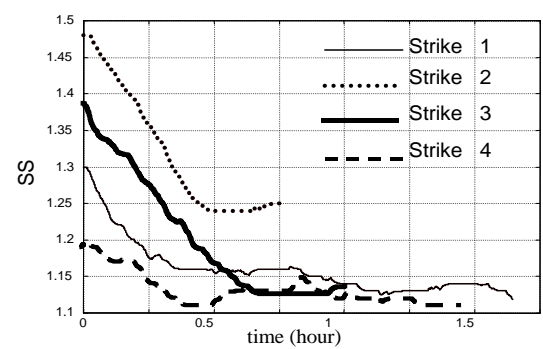

Fig. 4. Profiles of supersaturation during the boiling phase

The following figures show the results of the NLP technique for all the strikes. For each strike, the prediction on the mass of crystals is compared with the experimental observation $C_{obs}$.

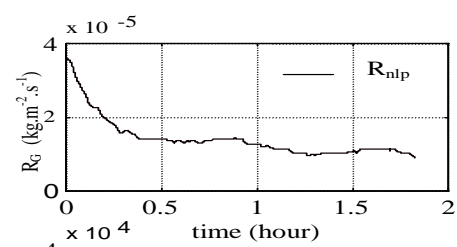

Fig. 8. Results of NLP technique for strike 4

### 5.2. Discussion

If estimates deduced from strike 1 are to be used in a MBPC approach as nominal parameters, it seems clear that they need some sort of on-line tuning to reach those of the other strikes. There is a need to adapt the model nonlinearities to those of the specific process at hand. This statement is close to the grey model conjecture postulated by Georgakis (1995). This postulate says that any fundamental nonlinear parameter-fitted model of a plant will need to be paired with a linear data-driven component if it is to be effectively used for on-line model predictive control. Following this idea, a linear correction on the order g should be sufficient here.

The example of the first empirical expression shows the difficulty of finding the proper parameterization for the growth rate model. For engineering practice, the classical model of growth rate seems the more appropriate. Nonetheless, these models offer poor predictive capability outside the narrow region where they were developed. It can be said that they perform local fitting.

Thus, another shade of grey was investigated in order to obtain a more flexible growth rate model.

## 6. Crystal growth rate model through hybrid ANN

Due to the complexity of the crystallization phenomena, and instead of using empirical expressions for the growth rate, Rawlings et al (1993) suggested the use of flexible nonlinear functions such as neural networks to model the growth rate as a direction of future work. Such flexible empiricism may prove to be useful in developing models with an intended control.

ANNs can be regarded as general purpose nonlinear functions for performing mappings between two sets of variables (Bishop, 1994). The mapping is governed by a set of parameters called weights whose values are determined on the basis of a set of examples of the variables. Furthermore, it can be demonstrated that a feedforward ANN with only one hidden layer is capable of approximating any continuous nonlinear function, provided the number of neurons is sufficiently large (Hornik et al., 1989). Indeed, ANNs are powerful tools where the theoretical knowledge is insufficient but experimental data available, as it is the case here.

### 6.1. Hybrid ANN

Fig. 5. Results of NLP technique for strike 1

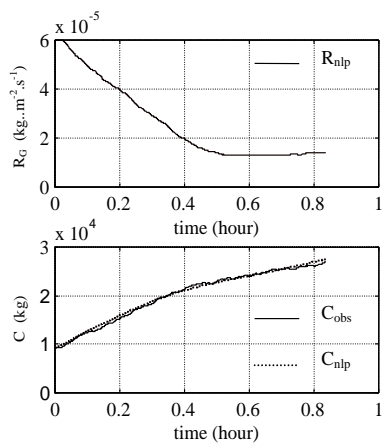

Fig. 6. Results of NLP technique for strike 2

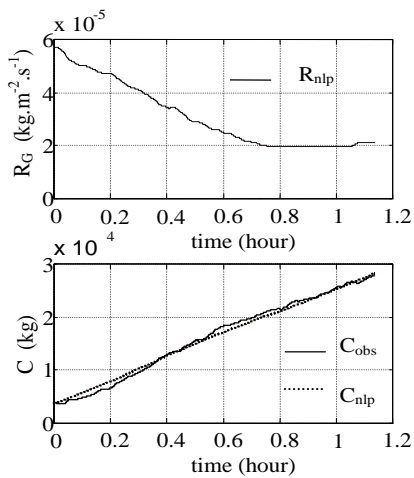

Fig. 7. Results of NLP technique for strike 3

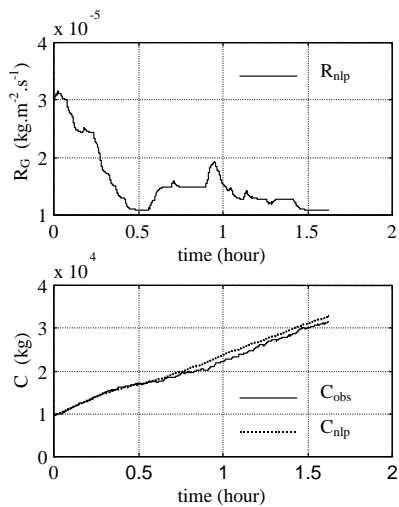

Most often, ANNs are used as black-box tools. Thus, there is no need to develop first principles models of the process since they have the ability to learn a solution from the presentation of a set of inputs and outputs issued from the process. The learning or training procedure consists of adjusting a network's set of internal parameters called weights, in order to minimize the squared error between the network's outputs and the desired outputs. Once the training stage is over, an ANN must generalize e.g. it must have capture the underlying trends in the training set in order to produce reliable outputs when it is presented with new input patterns (that have not been learned).

Nevertheless, it can be wasteful to throw away the a priori knowledge available on the process. A mixed modeling strategy that allows one to take advantage of the prior knowledge while retaining the flexibility of ANN is proposed here. This methodology introduces the concept of hybrid models (Psichogios & Ungar, 1992). The hybrid model is comprised of two parts. The first part consists of an ANN as an approximator of the growth rate. The second represents the physical knowledge formulated by the mass balance of the sucrose crystals. The network's output serves as an input to the mass balance ordinary differential equation (ODE). A sketch of the hybrid model is presented in Figure 9. The network architecture chosen here is a multilayer perceptron (MLP) with one hidden layer.

for updating the weights of the ANN component can be obtained by multiplying the observed error with the gradient of the hybrid model's output $C_{mod}$ with respect to the neural network's output $R_G$. In other words, the network's weights are changed proportionally to their effect on the mass of crystals predictions. The gradient $\chi$ can be calculated through integration of the sensitivity equation.

$$\frac{d\chi}{dt} = K_{AT}(C^{\frac{2}{3}} + \frac{2}{3}C^{-\frac{1}{3}}R_G\chi)$$  (5)

with :

$$\chi = \frac{dC}{dR_G}$$

and $K_{AT} = \dfrac{k_a.N^{\frac{1}{3}}}{(\rho \cdot k_v)^{\frac{2}{3}}}$

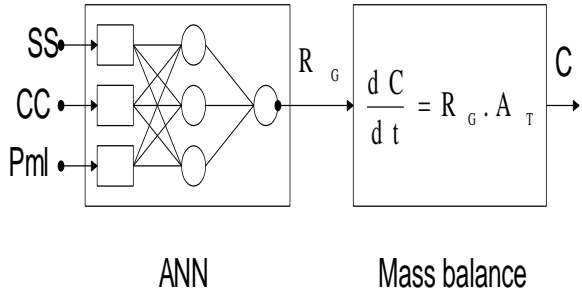

Fig. 9. Hybrid ANN

### 6.2. Hybrid training

The hybrid character leads to a specific training procedure. Actually, the training procedure of the ANN requires a knowledge of the growth rate, the target output in the neural network terminology. As its measurement is unavailable, the learning phase relies on the hybrid scheme. The observed error between the hybrid model's prediction and the observed mass of crystals are back-propagated through the mass balance equation and converted into an error signal for the neural network component. This hybrid training procedure is described in Figure 10. More precisely, a suitable error signal

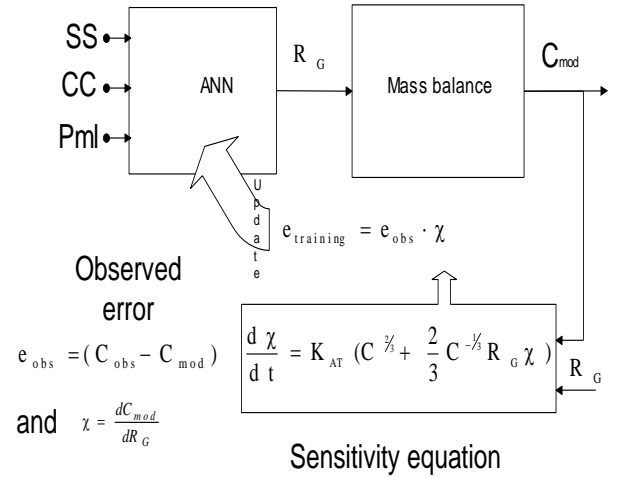

Fig. 10. Hybrid training procedure

The hybrid training was implemented through a modified version of the Levenberg-Marquartd training algorithm available under the MATLAB® neural network toolbox. The logistic sigmoïd activation function was chosen for all the neurons. As there was an order of magnitude between the dif-ferent variables, all experimental data were normalized between 0.1 and 0.9. This led to better performance in the training stage. The batch mode of learning was used e.g. the weights of the ANN were updated after a complete presentation of the profiles of a strike. More precisely, the profiles of supersaturation, crystal content and purity were together presented to the hybrid model. Their normalized values were propagated through the network part whose output gave an estimate of the growth rate. This estimation, after a denormalization, was propagated through the first-principles part of the hybrid model. As was the case for the NLP technique, the mass balance was integrated in order to obtain the model response. The observed error could be then calculated and converted into a suitable error signal for updating the network's weights by integrating the sensitivity equation. Several training sessions were carried out. The architecture with the best accuracy while avoiding overfitting, is a network with three hidden neurons. The network representation was obtained by using network pruning with an optimal brain surgeon (Bishop, 1995). This architecture results from a trade-off between the appropriate degree of flexibility of the nonlinear mapping and the non-fitting to the noise.

### 6.3. Results of the training phase

Strike 1, taken here as a reference, was used for training. $R_{ann}$ is the growth rate calculated by the hybrid ANN while $C_{ann}$ is the corresponding prediction on the mass of crystals.

brid ANN's prediction is close to that of the NLP technique ($R_{nlp}$). But at the same time, it is interesting to point out that the estimated growth rate profile possesses the appropriate degree of smoothness. Thus, overfitting does not occur.

### 6.4. Results of the generalization phase

Once the training was over, the resulting weights were frozen and tested on generalized strikes 2, 3 and 4. In the following figures, it can be seen that the closer $R_{ann}$ is to $R_{nlp}$, the better is the fit to $C_{obs}$. This fact is not surprising considering the previous NLP results. But there is a major difference here since these are the same network parameters that gave these predictions. To better appreciate this flexibility, kinetic parameters deduced from strike 1 (see line 1 of Table 7) are taken as nominal parameters in order to simulate the growth rates on the other strikes. It clearly appears that the simulated growth $R_1$ depends heavily on the local operating conditions. Consequently, the related mass predictions $C_1$ are far from the observed mass of crystals $C_{obs}$. In this comparison, strike 1 was taken as a base case since it was used for training. In other words, by making an analogy with the neural network context, the tested other strikes are simulated with the trained kinetic parameters $K_G$ and $g$ of strike 1.

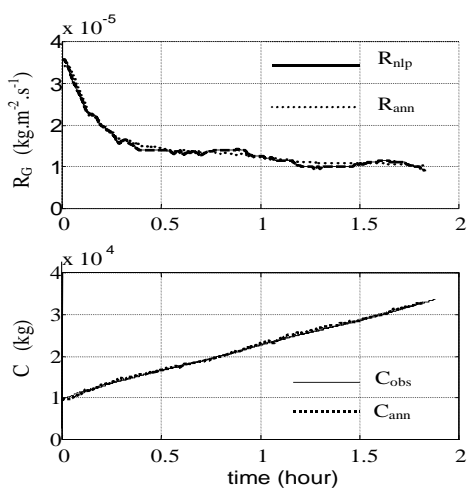

Fig. 11. Results of the training from strike 1

The hybrid model's prediction closely matches the observed mass of crystals. It is interesting to note that, as the network's output is constrained by the mass balance ODE, the estimated growth rate does not exhibit non-physical profiles and thus the hy-

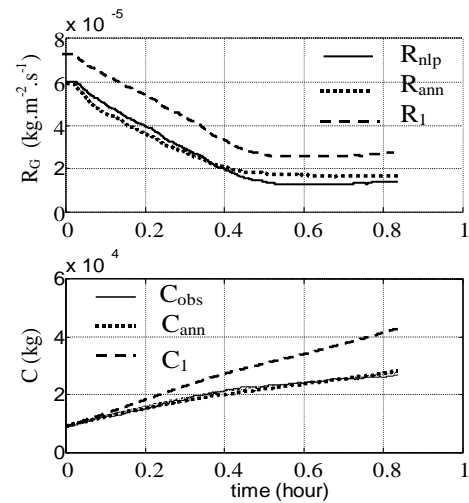

Fig. 12. Generalized strike 2

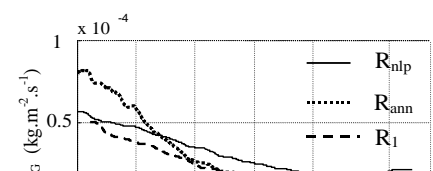

More discrepancies are observed for strike 3 at the beginning. Nonetheless, the gap to be filled is not as large as that from the NLP technique. Here, it seems possible to discard an adaptive procedure because the mismatch could be overcome by feedback compensation. Strike 1 was used for training. It was not a deliberate choice. Future work will be aimed at evaluating the hybrid ANN approach according different conditions. For example, training could be made on strike 2 (for which the profile of supersaturation is above the other ones, see Figure 4). It will be also interesting to test the hybrid ANN on a strike which exhibits very unusual profiles.

## 7. Conclusions

In order to develop a nonlinear model of the sugar crystallization process with an intended use in a MBPC scheme, two modeling approaches of the main process nonlinearity, the growth rate, were investigated.

The first approach was rather classical. It consisted of estimating the parameters of the empirical expressions through a NLP optimization technique. This modeling strategy imposes an a priori structure to the model. Two structures were studied. The first one contains four adjustable parameters and gives poor results, in that convergence to the optimum parameters was not guaranteed. The second structure, with two parameters, the most used in the crystallization engineering practice, overcomes the problem of convergence but does not resolve the problem of accuracy outside the region where the parameters were fitted.

The model performs local fitting. The choice of a correct parameterization may overcome these problems. But this strategy can be cumbersome as there is a limited knowledge on the crystallization phenomena.

The second approach is rather new. A structured model with two components was used to model the growth rate. The first component was an ANN as an approximator of the growth rate. The second component is the mass balance of the sucrose crystals. The latter represents the a priori knowledge and forces the output of the ANN to have physical sense.

This modeling strategy was termed hybrid.

The clear advantage of the ANN is not to impose an a priori structure to the model, thus reducing the modeling effort. Moreover, it appears that this second type of model, thanks to its greater flexibility, has better predictive capability under different operating conditions. In other words, this modeling technique is capable of expressing the process nonlinearities and complexities according to a general model structure.

The hybrid approach would be useful when there is a difficult-to-model process parameter included in a first-principles framework. This framework is

Fig. 13. Generalized strike 3

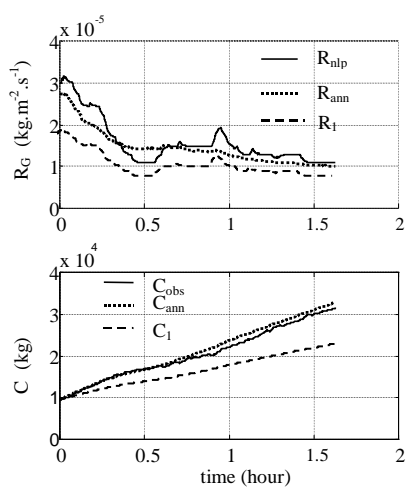

Fig. 14. Generalized strike 4

### 6.5. Discussion

Three interesting properties of ANN should be noted. First, ANN permits no a priori assumptions on the proper parameterization of the growth rate model. It is only necessary to specify the input space. Supersaturation, purity of mother liquor and crystal content were the natural candidates. Second, the parsimonious property of ANN (only three hidden neurons) leads to better generalization and more powerful computation, useful for a future real time implementation. The third is that the ANN are less sensitive to the signal's noise content than the NLP technique.

From the above figures, it is clear this second type of growth rate model offers a greater predictive capability under different operating conditions. Unlike the NLP technique, which models a limited number of nonlinearities, here a wider spectrum of nonlinearities are taken into account. Briefly, it extrapolates better. It is important to keep in mind that these are the same parameters or weights of the ANN that give the mass predictions of strikes 2 to 4.

generally represented by classical material and energy balances. The difficult-to-model parameter generally are rate terms (heat or mass transfer) with strong nonlinear effects, usually described by complex nonlinear expressions with a lot of adjustable parameters. It was demonstrated that the hybrid approach avoids the extensive experimentation needed to get the proper parameterization, and offers a more general and flexible model. Thus, the gain is double compared to classical approaches. The modeling effort is largely reduced and the model's predictions more accurate for various process conditions. Nonetheless, the main drawback of this modeling strategy is that training the ANN can be computationally intensive and time consuming. Indeed, the universal approximation theorems of ANN are not constructive in the sense that they are only existence theorems. But, computational effort does not mean modeling effort. After all, a good model needs some kind of effort and a good model gives good control because as accuracy increases so also does the performance of the controller. Furthermore, it can be said that, more and more, in the modern control techniques, the model becomes the controller.

## Acknowledgements

The authors gratefully acknowledge the technical staff of Bois Rouge sugar mill.

## Nomenclature

| | |
|---|---|
| $A_T$ [m$^2$] | total surface area of crystals |
| $k_a$ | surface shape factor |
| $k_v$ | volume shape factor |
| $k_g$ | growth rate constant |
| N | number of crystals |
| $K_{AT}$ | Constant associated with crystals surface area |
| $K_G$ | overall growth rate constant |
| g | growth rate order |
| e | exponent of slurry voidage to solid fraction |
| SS | Supersaturation |
| CC | Crystal content or solid fraction in the massecuite |
| Pml | Purity of mother liquor |
| $R_G$ [kg.m$^{-2}$.s$^{-1}$] | overall growth rate based on mass deposition |
| C [kg] | mass of crystals |

Subscripts

| | |
|---|---|
| ann | Growth and mass related to the hybrid ANN |
| nlp | Growth and mass related to the NLP technique |
| 1 | Growth simulated with parameter |

estimates issued from strike 1.

Greek letters

| | |
|---|---|
| $\alpha$ | coefficient for the impurity term |
| $\chi$ | gradient of hybrid model's output with respect to the growth rate |
| $\Theta$ | Vector of kinetic parameters |
| $\Theta_0$ | Initial guesses for the kinetic parameters |
| $\rho$ [kg.m$^{-3}$] | crystal density |

**Appendix**

The mathematical model of the sugar boiling process highlights the following state vector:

$$\mathbf{X} = \left[ W, S, I, C, T, \mu_1, \mu_2 \right]$$

W is the mass of water in the pan
S is the mass of sucrose in solution
I is the mass of impurities in solution
C is the mass of sucrose in crystal form
T is the temperature of massecuite
$\mu_1$ first moment of the distribution function
$\mu_2$ second moment of the distribution function

The following assumptions guides the design of the model:

- The pan is well-mixed thanks to a good circulation. Consequently, it is homogeneous in temperature and supersaturation
- The totality of the massecuite is in ebullition. The hydrostatic pressure effect is ignored. The temperature of massecuite is the ebullition temperature of water, derived from the vacuum in the pan , plus the boiling point elevation.
- Seeding takes place and the number of crystals is assumed to be constant all along the strike. This assumption leads to the following ones
- No primary nucleation, agglomeration nor breakage
- The Mac Cabe's law is verified e.g. the growth rate is independent of crystal size
- Growth rate dispersion does not occur

**Material balances**

$$\frac{dW}{dt} = x_e \, \overset{\bullet}{m_a} - \overset{\bullet}{m_{vap}}$$

$$\frac{dI}{dt} = x_{ns} \, \overset{\bullet}{m_a}$$

$$\frac{dS}{dt} = x_s \, \overset{\bullet}{m_a} - \frac{dC}{dt}$$

$$\frac{dC}{dt} = \rho k_v N \frac{d\mu_3}{dt} = 3\rho k_v NG\mu_2 = R_G N \mu_2 k_a = R_G A_T$$

$$R_G = 3\rho \frac{k_v}{k_a} G = 3 \frac{\rho}{F} G$$

$$F = \frac{k_a}{k_v}$$

$$x_s = \frac{B_x}{100} \cdot \frac{P}{100}$$

$$x_{ns} = \frac{B_x}{100} \cdot \frac{(100 - P)}{100}$$

$$x_e = \frac{(100 - B_x)}{100}$$

$\overset{\bullet}{m_a}$ feed rate of syrup to the pan(kg.s$^{-1}$)
$\overset{\bullet}{m_{vap}}$ evaporation rate of water from the pan (kg.s$^{-1}$)
$\rho$ density of sucrose crystals(kg.m$^{-3}$)
$N$ number of crystals
$k_v$ volume shape factor, $k_a$ surface shape factor
F shape factor
$\mu_3$ 3$^{th}$ moment of the crystal size distribution function(m$^3$)
$B_x$ Brix of feed syrup (concentration of dissolved solids in feed)
P Purity of feed syrup ( fraction of sucrose in solids of feed)
$x_e, x_s, x_{ns}$ mass fractions of water, sucrose and impurities in the feed syrup
G overall linear growth rate (m.s$^{-1}$)
$R_G$ overall growth rate based on mass deposition (kg.m$^{-2}.s^{-1}$)

**Energy balance**

$$\frac{d(C_p MT)}{dt} = \overset{\bullet}{m_a} \, h_a - \overset{\bullet}{m_{vap}} \, h_{vap} + \overset{\bullet}{Q} + \overset{\bullet}{W} + \frac{dC}{dt} L_{ls}$$

$C_p$ specific heating capacity of massecuite (J.kg$^{-1}$.($^{\circ}C$)$^{-1}$)
$M$ mass of the massecuite (kg)
T temperature of the massecuite ($^{\circ}C$)
$h_a$ specific enthalpy of feed syrup (J.kg$^{-1}$)
$h_{vap}$ vaporization enthalpy (J.kg$^{-1}$)
$L_{ls}$ crystallization enthalpy (J.kg$^{-1}$)
$\overset{\bullet}{Q}$ heat input (J.s$^{-1}$)
$\overset{\bullet}{W}$ stirrer power (J.s$^{-1}$)

**Heat transfer**

$$\overset{\bullet}{Q} = UA(T_v - T) = \overset{\bullet}{m_v} L_{lv}$$

$U$  overall heat transfer coefficient  $(W.m^{-2}.(°C)^{-1})$

$A$  total area of heat transfer $(m^2)$

$T_v$  steam temperature $(°C)$

$T$  massecuite temperature  $(°C)$

$\overset{\bullet}{m_v}$  steam flowrate (kg.s$^{-1}$)

$L_{lv}$  latent heat of  condensation (J.kg$^{-1}$)

## Supersaturation and solubility

Supersaturation SS :

$$SS = \frac{100 - SAT}{SAT} \times \frac{S}{W} \times \frac{1}{SC}$$

Solubilty correlation :

$$SAT = 64,407 + 0,0725 \times T + 0,002057 \times T^2 - 9,035 \times 10^{-6} \times T^3$$

Solubility coefficient (Wright,1974):

$$SC = 1.0 - 0.088 \frac{I}{W}$$

## Population balance

$$\frac{1}{V} \frac{\partial(nV)}{\partial t} + G \frac{\partial(n)}{\partial L} = 0$$

$n$  population density  $(m^{-4})$

$V$  volume of the massecuite $(m^3)$

According to the assumptions, this balance can be rewritten with the normalized distribution  function$f(L,t)$ ,

$$\frac{\partial f}{\partial t} + G \frac{\partial f}{\partial L} = 0$$

$f$  ( m$^{-1}$ )

## Moments of the crystal size distribution

$$\frac{d(\mu j)}{dt} - j \int_0^\infty L^{j-1} G(L,t) f(L,t) dL = 0$$

if G is independant of L (Mac Cabe's law),

$$\frac{d(\mu_j)}{dt} - jG\mu_{j-1} = 0$$

According to the assumptions (no nucleation nor growth rate dispersion) ,

$$\frac{d(\mu_0)}{dt} = 0$$

$$\frac{d(\mu_1)}{dt} = G$$

$$\frac{d(\mu_2)}{dt} = 2G\mu_1$$

$$\frac{d(\mu_3)}{dt} = 3G\mu_2$$

$\mu_0, \mu_1(m), \mu_2(m^2), \mu_3(m^3)$ moments of the crystal size distribution